\documentclass[5pt]{article}

\usepackage[letterpaper]{geometry}
\usepackage{spconf,amsmath,epsfig}
\usepackage{enumitem}

\usepackage{fancyhdr}
\begin{document}\sloppy

% Example definitions.
% --------------------
\def\x{{\mathbf x}}
\def\L{{\cal L}}

% Title.
% ------
\title{Co-projection-plane based 3-D padding for polyhedron projection for 360-degree video}
%
% Single address.
% ---------------
\name{Li Li$^{\ast}$, Zhu Li$^{\ast}$, Xiang Ma$^{\star}$, Haitao Yang$^{\star}$ and Houqiang Li$^{\dagger}$\thanks{This paper is partially supported by the UMKC strategic funding on big imaging and smart city center}}
\address{$^{\ast}$ University of Missouri Kansas City \\
    $^{\star}$ Huawei Technologies Co., Ltd. \\
    $^{\dagger}$ University of Science and Technology of China\\
    \{lil1, lizhu\}@umkc.edu, \{maxiang6, haitao.yang\}@huawei.com, lihq@ustc.edu.cn}

\maketitle

\begin{abstract}
The polyhedron projection for 360-degree video is becoming more and more popular since it can lead to much less geometry distortion compared with the equirectangular projection. 
However, in the polyhedron projection, we can observe very obvious texture discontinuity in the area near the face boundary. 
Such a texture discontinuity may lead to serious quality degradation when motion compensation crosses the discontinuous face boundary. 
To solve this problem, in this paper, we first propose to fill the corresponding neighboring faces in the suitable positions as the extension of the current face to keep approximated texture continuity. 
Then a co-projection-plane based 3-D padding method is proposed to project the reference pixels in the neighboring face to the current face to guarantee exact texture continuity. 
Under the proposed scheme, the reference pixel is always projected to the same plane with the current pixel when performing motion compensation so that the texture discontinuity problem can be solved. 
The proposed scheme is implemented in the reference software of High Efficiency Video Coding. 
Compared with the existing method, the proposed algorithm can significantly improve the rate-distortion performance. 
The experimental results obviously demonstrate that the texture discontinuity in the face boundary can be well handled by the proposed algorithm.
\end{abstract}
\section{Introduction}
\label{sec:intro}
% 360-degree video brief intro
Along with the emergence and popularity of one virtual reality (VR) product after another, such as Oculus Rift, Gear VR, and HTC Vive, video contents are becoming one of the most important applications for the VR product.
To support the content representation from all directions and create a fully immersed experience, the VR video needs to contain the information from all 360 degrees.
Therefore, the VR video, also named as 360-degree video, should be with very high spatial resolution even higher than $8K$ to maintain relatively good visual quality.
Such high resolution videos can bring many challenges to the video compression technologies, and the need to develop specified compression method for these video becomes quite urgent.

\begin{figure}[t]
  \centering
  \centerline{\includegraphics[width=8.0cm]{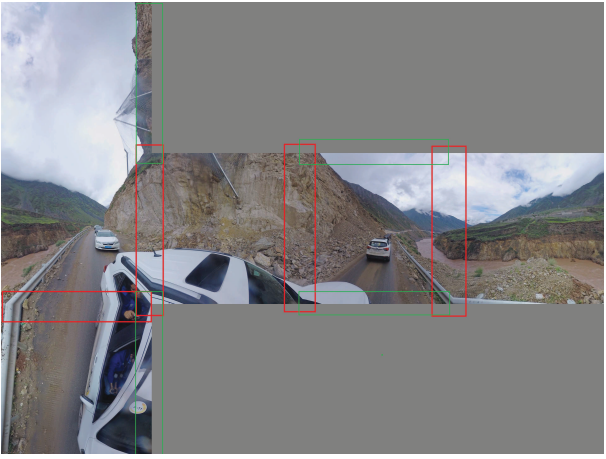}}
\caption{Typical example of texture discontinuity}
\label{fig:example}
\end{figure}

% projection, cubic format
Since the original 360-degree video is a sphere, to adapt to the modern video coding standards such as H.264/Advanced Video Coding (AVC) \cite{Wiegand2003}, and H.265/High Efficiency Video Coding (HEVC) \cite{Sullivan2012}, the 360-degree video is always projected to a 2-D format for compression.
According to the investigation in \cite{He2016}, there are actually lots of projection methods such as equirectangular and polyhedron including cube map, octahedron, icosahedron.
Comparing the equirectangular and polyhedron formats, the polyhedron formats present less geometry distortion so that they can lead to better coding efficiency \cite{Zhou2016} \cite{Zhou20162}.
However, the polyhedron formats also have their disadvantages that very obvious texture discontinuities exist in the area near the face boundary.
The texture discontinuities can be divided into two kinds, which are obviously shown in Fig.~\ref{fig:example} for the typical $4\times3$ cubic format.
One kind of the discontinuities is caused by the face unfold from 3-D cubic to 2-D image, which is represented by the green rectangles.
The other kind of discontinuities is brought by the projection to different planes (or faces) from sphere to cubic format, which is shown by the red rectangles.
When the motion vector (MV) happens to cross the face boundary, the current motion compensation (MC) scheme will obtain an unreasonable prediction block with quite obvious texture discontinuity, which will lead to serious coding efficiency decrease.

In the current standard-based video coding scheme, a simple padding scheme, which extends the picture boundary pixel to the outside of the picture, is implemented in the HEVC reference software \cite{HM16.6} to both guarantee the picture size as the multiple of the coding unit size and prevent the MC operation from crossing the picture boundary.
Li \emph{et al.} \cite{Li2010} have also tried to optimize the padding scheme for arbitrary size picture using the fundamental rate distortion optimization (RDO) theory. 
However, since these schemes only consider the picture itself and have not considered the specific 360-degree information of the 360-degree video, they are not the best ways to solve the problems of texture discontinuity in the face boundary for the 360-degree video.

% contribution
Therefore, in this paper, to better solve the problem of texture discontinuity in the face boundary, we try to make full use of all the information from the 360-degree video.
To be more specific, we first fill the neighboring faces in the suitable positions for the current face to keep approximate texture continuity.
Then we propose a co-projection-plane based 3-D padding method to project the reference pixels in the neighboring face to the current face to guarantee exact texture continuity.
Under the proposed scheme, the reference pixel is always projected to the same plane with the current pixel when performing MC so that the texture discontinuity problem in the face boundary can be solved.

% paper organization
This paper is organized as follows.
In Section \ref{sec:cube_map}, we will give a brief introduction of the polyhedron projection.
The proposed co-projection-plain based 3-D mapping method will be described in detail in Section \ref{sec:proposed_algorithm}.
The detailed experimental results will be shown in Section \ref{sec:experiments}.
Section \ref{sec:conclusion} concludes the whole paper.

\section{A brief introduction of the polyhedron projection}
\label{sec:cube_map}
\begin{figure}[t]
  \centering
  \centerline{\includegraphics[width=4.0cm]{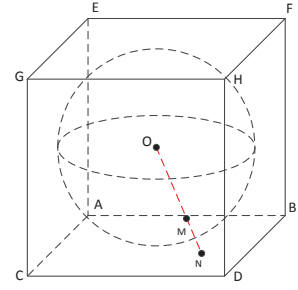}}
\caption{Cube map projection from inscribed sphere}
\label{fig:cube_map_projection}
\end{figure}

As its name implies, polyhedron projection is to project the inscribed sphere (360-degree video) to each face of the polyhedron, such as cube, octahedron, and icosahedron.
As a typical example, the detailed projection process from inscribed sphere to the cube map can be seen from Fig.~\ref{fig:cube_map_projection}.
For each point $N$ in the face of the cube, we will connect a line between the center point $O$ and $N$.
Then the line and the sphere will have an intersection point $M$, and the pixel value of point $M$ will be used as the value of point $N$.
Since the point $M$ may not be in the integer sampling position of the sphere, the pixel value of point $M$ will be interpolated through surrounding integer pixels.
To be more specific, the Luma component is interpolated using the Lanczos3 ($6\times6$) \cite{Lanczos} interpolation filter, and the Chroma component is interpolated using the Lanczos2 ($4\times4$) \cite{Lanczos} interpolation filter.

\begin{figure}[t]
\begin{minipage}[b]{0.48\linewidth}
  \centering
  \centerline{\includegraphics[width=3.8cm]{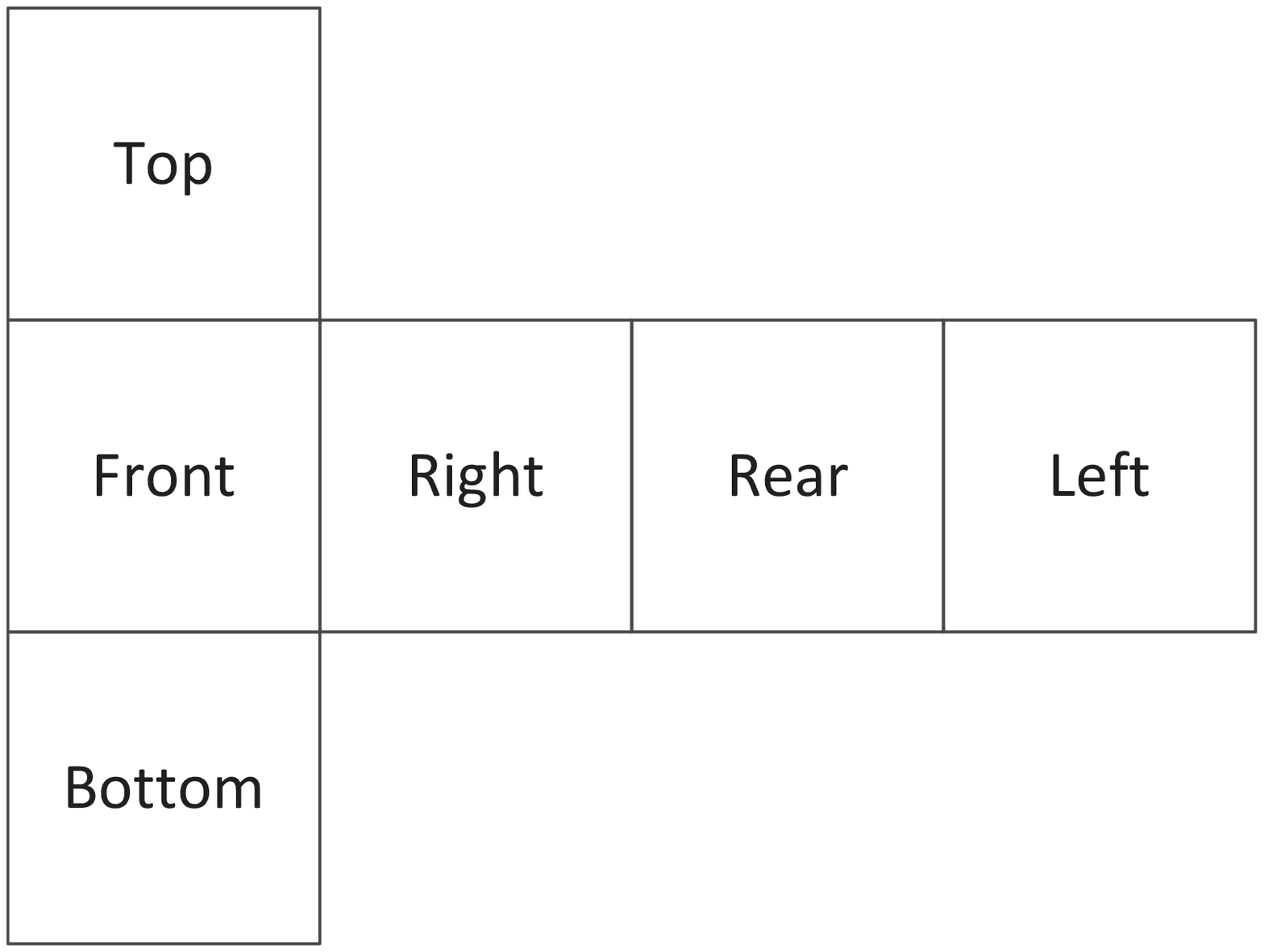}}
  \vspace{0.5cm}
  \centerline{(a) $4\times3$}\medskip
\end{minipage}
\begin{minipage}[b]{0.48\linewidth}
  \centering
  \centerline{\includegraphics[width=3.0cm]{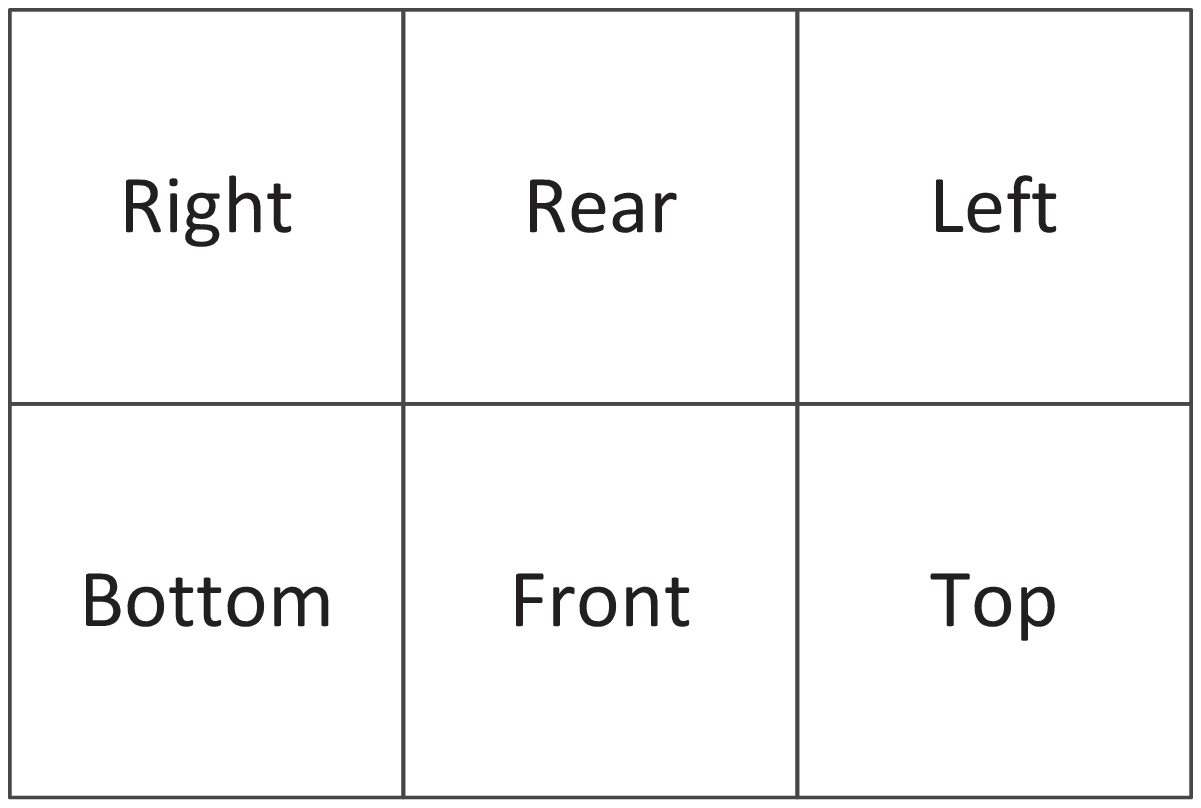}}
  \vspace{1.0cm}
  \centerline{(b) $3\times2$}\medskip
\end{minipage}
\caption{Typical unfold cubic format}
\label{fig:unfold}
\end{figure}

After the projection from a sphere to a polyhedron, the polyhedron will then be unfolded to obtain the 2-D image for compression.
There are various kinds of unfolding methods for a polyhedron including non-compact and compact methods.
Especially, for the cube map projection, as shown in Fig. \ref{fig:unfold}, mainly two methods of unfolding by putting different faces in different positions are introduced, including $4\times3$ and $3\times2$ formates.
And in the following sections, the $4\times3$ cube map projection will be used as an example to introduce the proposed co-projection-plain based 3-D mapping methods.

\section{The proposed co-projection-plain based 3-D padding}
\label{sec:proposed_algorithm}
The proposed co-projection-plain based 3-D padding method will be introduced in two aspects.
We will first fill the corresponding neighboring faces in the suitable positions as the extension of the current face to keep approximated texture continuity in subsection \ref{subsec:approximated}.
Then we will project the reference pixels in the neighboring face to the current face to guarantee exact texture continuity in subsection \ref{subsec:exact}.
Finally, in subsection \ref{subsec:implementation}, we will introduce some implementation details.

\subsection{Approximated texture continuity}
\label{subsec:approximated}
\begin{figure}[t]
\begin{minipage}[b]{0.48\linewidth}
  \centering
  \centerline{\includegraphics[width=3.8cm]{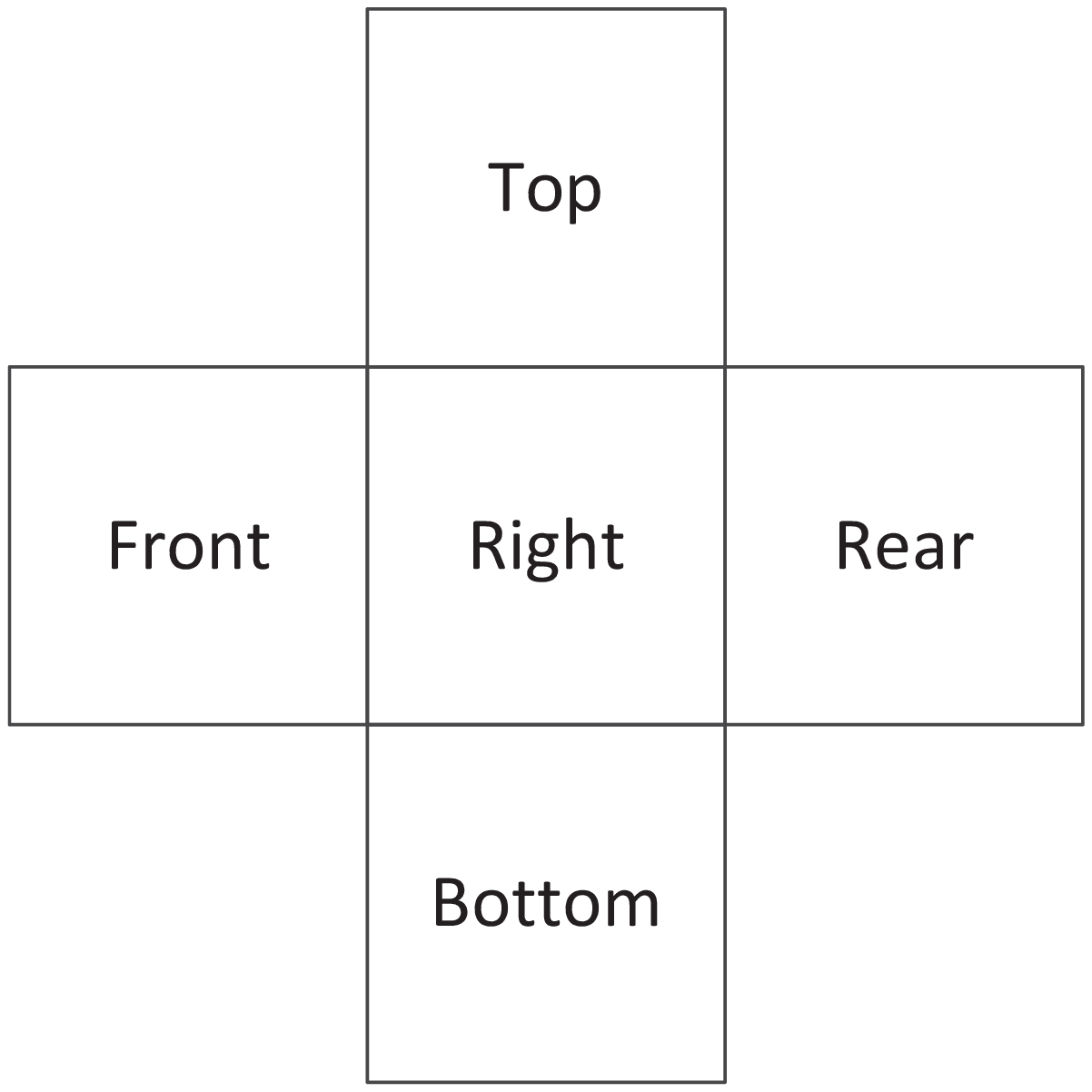}}
  \vspace{0.5cm}
  \centerline{(a) Model complementation}\medskip
\end{minipage}
\begin{minipage}[b]{0.48\linewidth}
  \centering
  \centerline{\includegraphics[width=3.8cm]{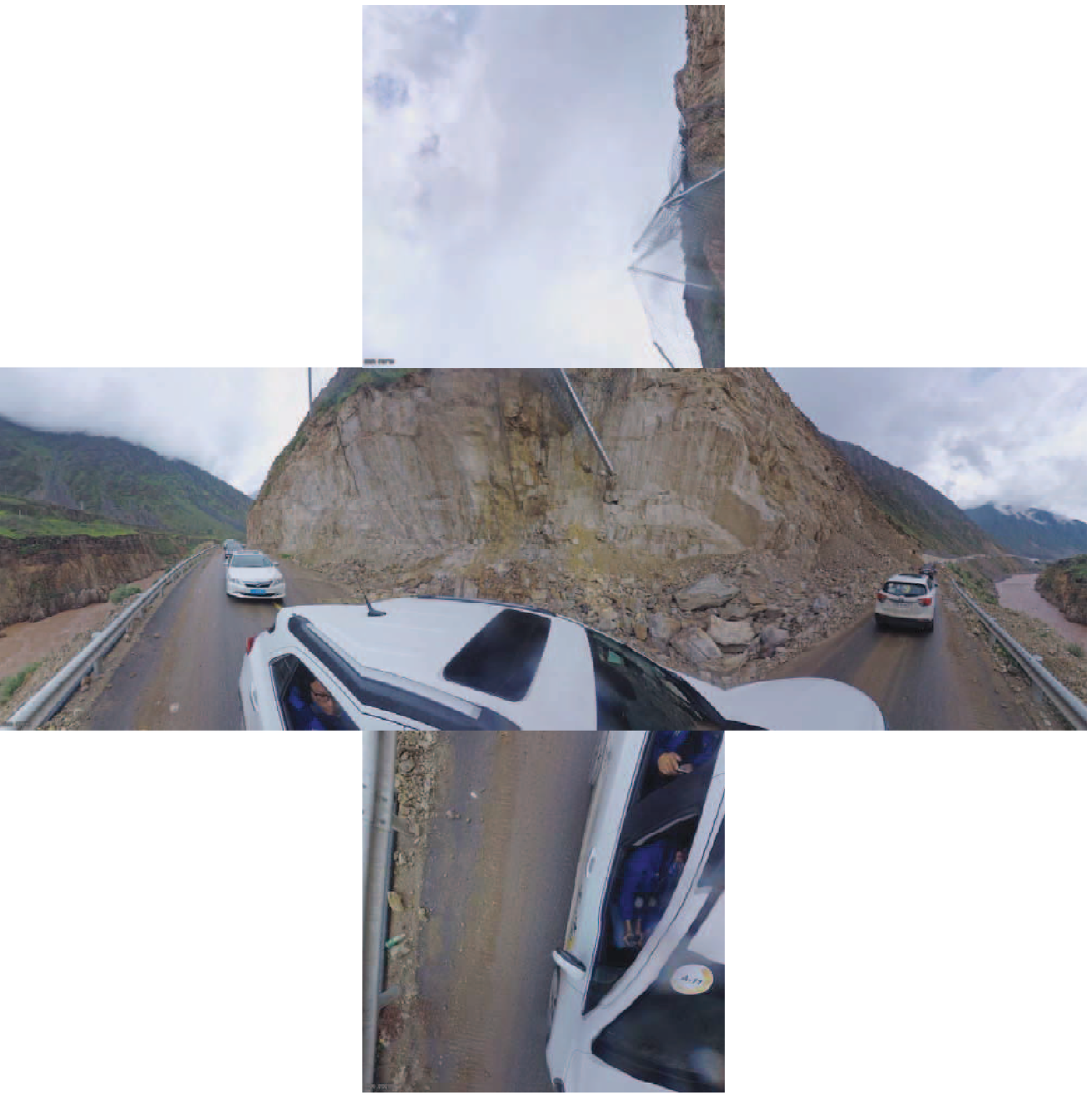}}
  \vspace{0.5cm}
  \centerline{(b) Actual complementation}\medskip
\end{minipage}
\caption{Typical complementation results}
\label{fig:complement}
\end{figure}

As each face of a cube has four edges, to achieve approximated texture continuity, we should first try to make all the four neighboring faces of the current face available.
As shown in Fig. \ref{fig:unfold} (a), the front face has three neighboring faces, the right and rear faces have two neighboring faces, and the top, bottom, and left faces have only one neighboring face.
We will complement the neighboring faces of all the faces to four neighboring faces.
Using the right face as an example, besides the existing front and rear faces, we will complement the top and bottom faces for the current face.
The complementation result is shown in Fig. \ref{fig:complement} (a), and the actual result of a typical sequence is presented in Fig. \ref{fig:complement} (b).

\begin{figure}[t]
\begin{minipage}[b]{0.48\linewidth}
  \centering
  \centerline{\includegraphics[width=3.8cm]{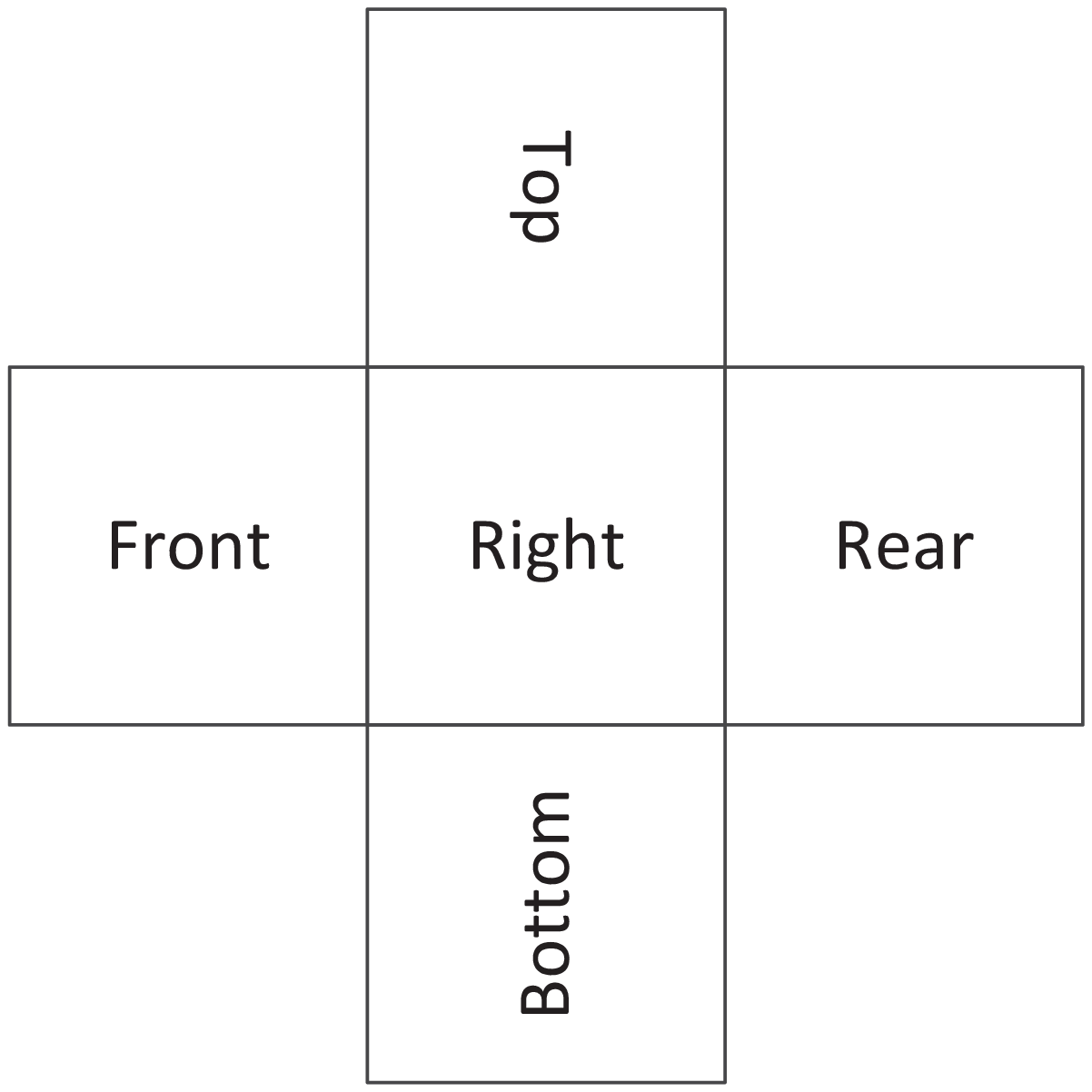}}
  \vspace{0.5cm}
  \centerline{(a) Model complementation}\medskip
\end{minipage}
\begin{minipage}[b]{0.48\linewidth}
  \centering
  \centerline{\includegraphics[width=3.8cm]{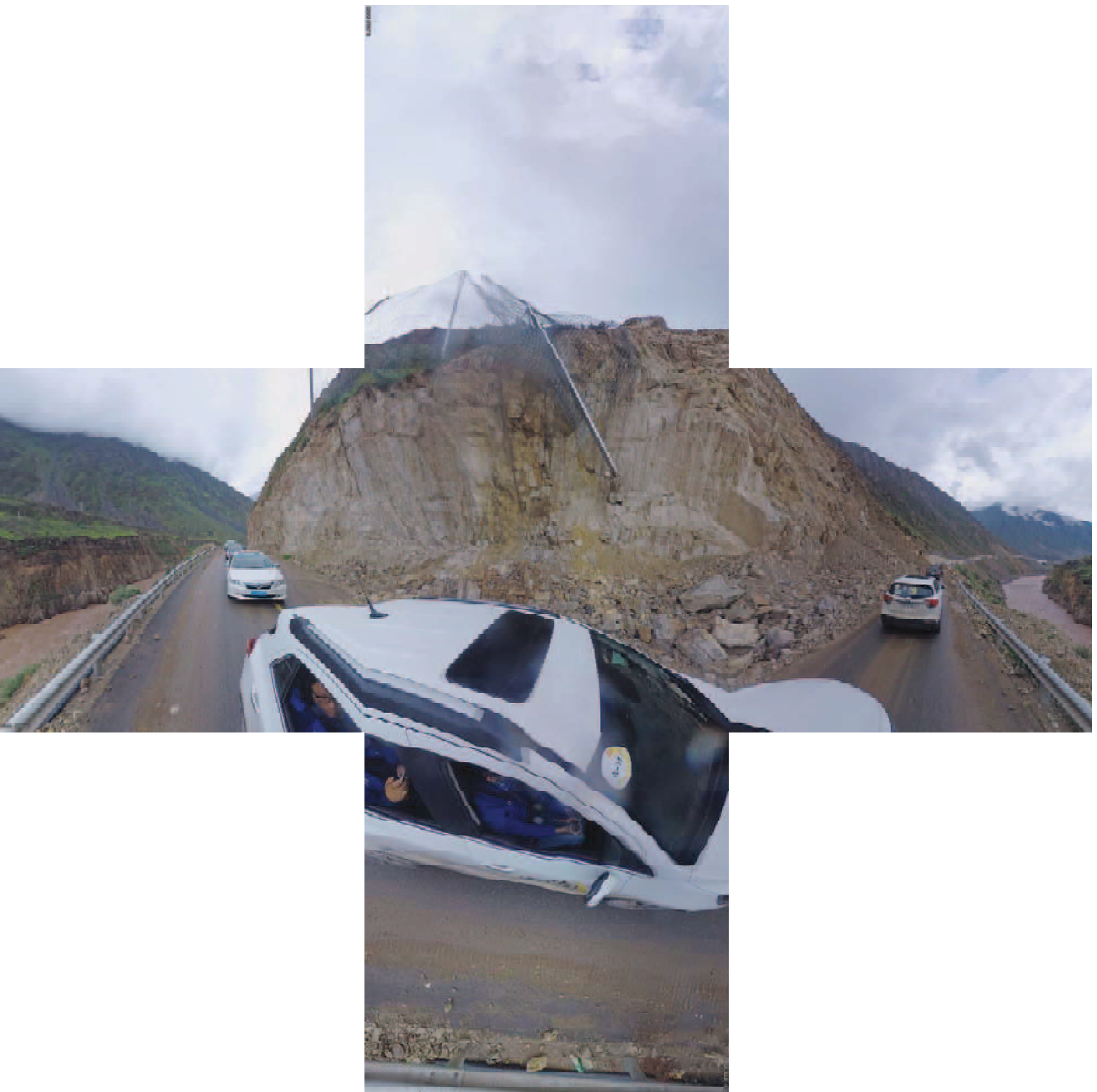}}
  \vspace{0.5cm}
  \centerline{(b) Actual complementation}\medskip
\end{minipage}
\caption{Typical approximated texture continuity}
\label{fig:approximate}
\end{figure}

As can be obviously seen from Fig. \ref{fig:complement} (b), the complementation result still presents very obvious texture discontinuity in the common edges between the center face and top/bottom faces.
The main reason is that the common edges of the neighboring faces are not aligned together.
To guarantee the alignment of the common edges, the top face should be rotated by 90 degrees clockwise, and the bottom face should be rotated by 90 degrees anticlockwise.
The final approximated texture continuity results are shown in Fig. \ref{fig:approximate}.
The above process is just a typical example for the right face, and the other faces can be done in a similar way to achieve approximated texture continuity.

\subsection{Exact texture continuity}
\label{subsec:exact}
After the approximated texture continuity is achieved, if we take a look at Fig. \ref{fig:approximate} (b) carefully, we can still see that straight lines on the car become broken lines when crossing the face boundary.
This is mainly caused by the cube map projection from inscribed sphere to difference faces.
Therefore, in this subsection, we will propose a co-projection-plain based 3-D padding to achieve exact texture continuity.

\begin{figure}[t]
  \centering
  \centerline{\includegraphics[width=8.0cm]{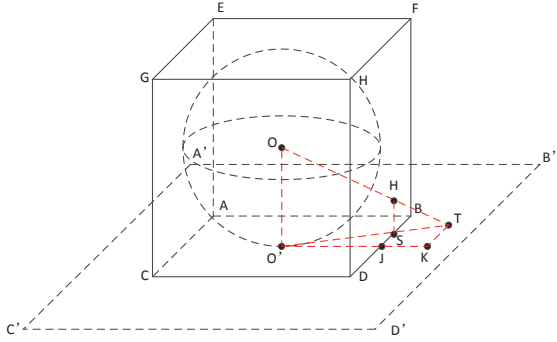}}
\caption{Co-projection-plain based 3-D padding}
\label{fig:projection}
\end{figure}

As shown in Fig. \ref{fig:projection}, under the co-projection-plain based 3-D padding method, we will try to extend the current face $ABCD$ into a larger one $A'B'C'D'$, and the values of the extended pixels will be determined by the projection of the neighboring faces, which are generated in the approximated-texture-continuity step, to the current face.
Using the bottom face as an example, for a point $T$ in the extended zone of the bottom face, assume that the top left position $A'$ is $(0,0)$, the position $T$ in the extension face is $(x,y)$, the face extension range is $S$, and the edge length of the cube is $a$.
Then the lengths of $TK$ and $JK$ can be calculated as
\begin{equation}
\label{eq:TK}
TK = \frac{a}{2}+S-y
\end{equation}
\begin{equation}
\label{eq:JK}
JK = x-a-S
\end{equation}
Therefore, according to the principle of similar triangles, we can obtain the length of $HS$ as
\begin{equation}
\label{eq:x}
HS = \frac{ST}{O'T} \times OO' = \frac{JK}{O'K} \times OO'
\end{equation}
Similarly, we can also obtain the length of $SJ$ as
\begin{equation}
\label{eq:y}
SJ = \frac{O'J}{O'K} \times TK
\end{equation}
In this way, the coordinate of the corresponding position in the right face can be derived.
The other projection positions of the neighboring faces can be derived in a similar way.

\begin{figure}[t]
\begin{minipage}[b]{0.48\linewidth}
  \centering
  \centerline{\includegraphics[width=3.8cm]{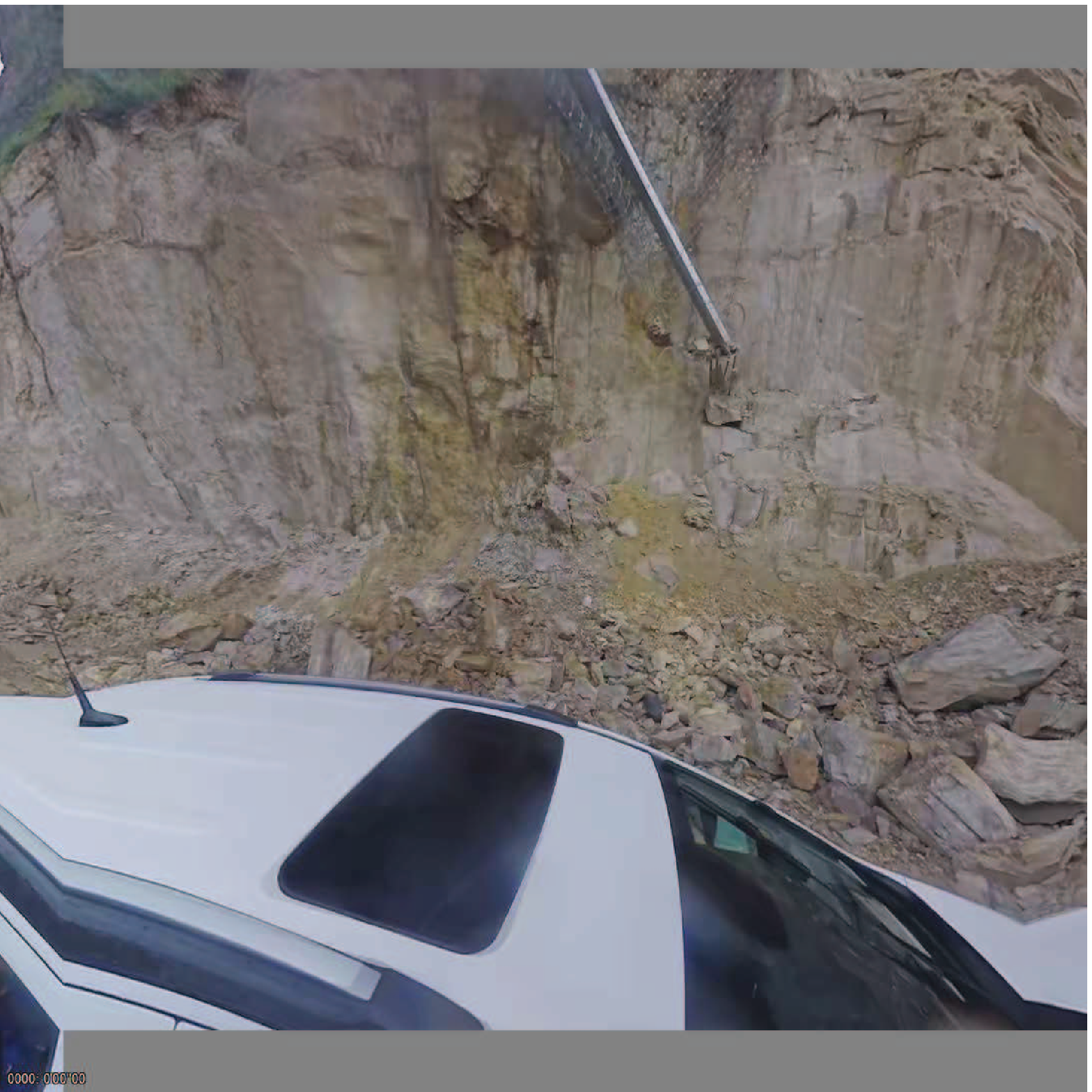}}
  \vspace{0.5cm}
  \centerline{(a) Original face extension}\medskip
\end{minipage}
\begin{minipage}[b]{0.48\linewidth}
  \centering
  \centerline{\includegraphics[width=3.8cm]{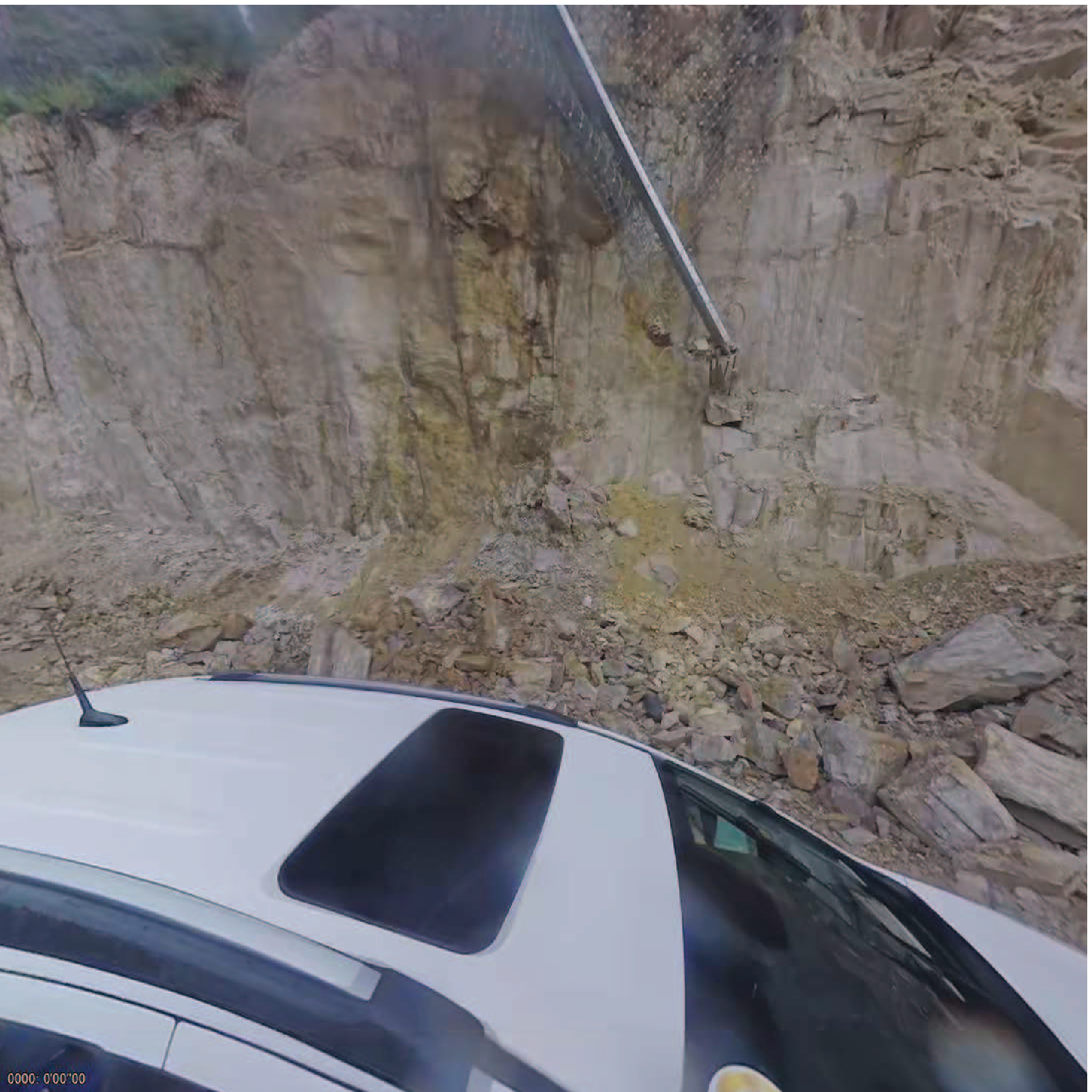}}
  \vspace{0.5cm}
  \centerline{(b) Proposed face extension}\medskip
\end{minipage}
\caption{Face extension comparison}
\label{fig:extension}
\end{figure}

It should be noted that the calculated coordinate may not be always in the integer position.
In the current implementation, the bilinear interpolation is used to interpolate the pixels in the fractional positions.
It should also be mentioned that the pixels belonging to lines $AA'$, $BB'$, $CC'$, and $DD'$ will be projected to the common edges of two neighboring faces.
If the bilinear interpolation is still used, the final pixel values will be interpolated from the neighboring pixels coming from two different faces, which is obviously unreasonable.
In our implementation, the pixels belonging to lines $AA'$, $BB'$, $CC'$, and $DD'$ are derived through the average of the neighboring pixels in the extended zones.
After these operations, the interpolation results are shown in Fig. \ref{fig:extension} (b).
Compared with the results generated by the HEVC reference software as shown in Fig. \ref{fig:extension} (a), it can be obviously seen that the proposed algorithm can achieve exact texture continuity.
Not only the gray zones but also the discontinuous face boundaries are filled with suitable values to guarantee exact texture continuity.

\subsection{Implementation details}
\label{subsec:implementation}
The proposed algorithm is implemented in the HEVC reference software.
Our current implementation can be roughly divided into two parts and will not lead to any modification of the coding tools in the coding unit (CU) level.
The first part is to get the extension for all the $6$ faces for the reference frames.
To be more specific, after the encoding of the current frame is finished, if the current frame is a reference frame, the neighbor faces of all the $6$ faces will be first complemented using the method introduced in subsection \ref{subsec:approximated} to generate the image similar to Fig. \ref{fig:approximate} (b).
Then the method introduced in subsection \ref{subsec:exact} will be used to generate the extended faces similar to Fig. \ref{fig:extension} (b) to achieve exact texture continuity.

\begin{figure}[t]
  \centering
  \centerline{\includegraphics[width=8.0cm]{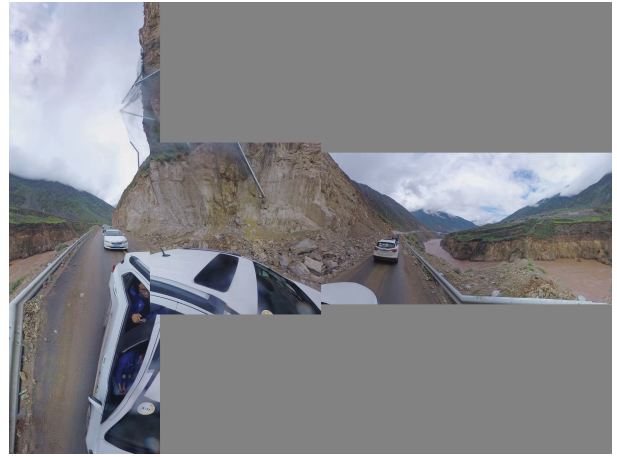}}
\caption{Typical reference frame}
\label{fig:reference}
\end{figure}

Then the second part is to fill the reference frame with the face extension when encoding each CU.
For example, when we are encoding a CU in the right face, we will fill in the right face extension to the each reference frame for the current CU.
The results can be seen from Fig. \ref{fig:reference}.
It seems discontinuous for the whole frame but for the right face in a predefined search range $S$, the texture is continuous.
And after the coding of CUs belonging to the current face, the reference frame will be refilled with the original values and prepare to be filled with the extension of other faces in the future encoding process.
It should be noted that in the decoding process before the reference frame will be used for each CU, we will already know the MV of the current CU.
Therefore, we can determine whether the current CU needs to fill in the extension of a current face or not according to the value of MV so as to avoiding the unnecessary extension operations and reducing decoding complexity.

\section{Experimental results}
\label{sec:experiments}
The proposed co-projection-plain based 3-D padding method is implemented in the HEVC reference software HM-16.6 to compare with HEVC without the proposed algorithm.
All the test conditions specified for inter frames including random access (RA) main $10$, low delay (LD) main $10$, low delay P (LDP) main $10$ are used as the test conditions.
The quantization parameters (QP) tested in our experiments are 22, 27, 32, 37 following the HEVC common test conditions.
The face extension range $S$ is set as $64$ in our experiments.
Besides, the BD-rate (Bjontegaard Delta rate) \cite{Bjontegaard2001} is used to measure the difference between the anchor and the proposed algorithm.
In the current implementation, the Peak Signal to Noise Ratio (PSNR) is used to measure the quality of between the reconstructed and original sequences.
We will use the quality metrics, which are more suitable for 360-degree videos such as WS-PSNR \cite{Sun2016} and S-PSNR \cite{Yu2015}, as the quality measurements in our future work.

For the test sequences, we use the test sequences specified in \cite{Boyce2016} to measure the performance of the proposed algorithm.
To be more specific, we used the conversion tool specified in \cite{He2016} to convert the high fidelity input test sequences in equirectangular format to the $10$ bit $4\times3$ cubic formate test sequences.
The detailed information and characteristics of the test sequences can be seen in Fig. \ref{tab:seq}.
The frame count tested is approximated as $1$ second as shown in Fig. \ref{tab:seq}.

\begin{table}[t]
\begin{center}
\caption{The characteristics of the test sequences} \label{tab:seq}
\begin{tabular}{|c|c|c|}
  \hline
  % after \\: \hline or \cline{col1-col2} \cline{col3-col4} ...
Sequence name & Resolution & frame count
  \\
  \hline
  Train\_le                & $4736\times3552$ & 64 \\
  SkateBoardingTrick\_le   & $4736\times3552$ & 64 \\
  SkateboardInLot          & $4736\times3552$ & 32 \\
  ChairLift                & $4736\times3552$ & 32 \\
  KiteFlite                & $4736\times3552$ & 32 \\
  Harbor                   & $4736\times3552$ & 32 \\
  PoleVault\_le            & $3840\times2880$ & 32 \\
  AerialCity               & $3840\times2880$ & 32 \\
  DrivingInCity            & $3840\times2880$ & 32 \\
  DrivingInCountry         & $3840\times2880$ & 32 \\
  \hline
\end{tabular}
\end{center}
\end{table}

The test results of the proposed algorithm in RA main10, LD main10, and LDP main10 are shown in Table \ref{tab:RA}, Table \ref{tab:LD}, and Table \ref{tab:LDP}, respectively.
From the test results, we can see that about for the Y component, compared with the HEVC anchor, about averagely $1.1\%$, $1.2\%$ and $1.2\%$ R-D performance improvement can be achieved in RA, LD, and LDP cases, respectively.
For U and V components, about averagely $1.3\%$, $1.5\%$, and $1.3\%$ bitrate reduction are observed accordingly.
Besides, we can also see from these tables that for the sequence with relatively larger motion, the maximum bitrate saving for the Y component can be as high as $3.3\%$, $3.4\%$, and $3.3\%$ in RA, LD, and LDP cases, respectively.

\begin{table}[t]
\begin{center}
\caption{The performance in RA case} \label{tab:RA}
\begin{tabular}{|c|c|c|c|}
  \hline
  % after \\: \hline or \cline{col1-col2} \cline{col3-col4} ...
Sequence name & Y & U & V               
  \\
  \hline
  Train\_le                & --0.2\%  &  --0.1\%  & --0.1\% \\
  SkateBoardingTrick\_le   & --0.4\%  &  --0.9\%  & --0.7\% \\
  SkateboardInLot          & --0.9\%  &  --1.2\%  & --2.5\% \\
  ChairLift                & --2.6\%  &  --3.2\%  & --3.0\% \\
  KiteFlite                & --0.1\%  &  --0.1\%  & --0.1\% \\
  Harbor                   & --0.1\%  &  --0.8\%  & --0.3\%   \\
  PoleVault\_le            & --0.2\%  &  --0.1\%  & --0.2\% \\
  AerialCity               & --2.1\%  &  --2.1\%  & --1.8\% \\
  DrivingInCity            & --0.6\%  &  --1.0\%  & --1.0\% \\
  DrivingInCountry         & --3.3\%  &  --3.6\%  & --3.3\% \\
  average                  & --1.1\%  &  --1.3\%  & --1.3\% \\
  \hline
\end{tabular}
\end{center}
\end{table}

\begin{table}[t]
\begin{center}
\caption{The performance in LD case} \label{tab:LD}
\begin{tabular}{|c|c|c|c|}
  \hline
  % after \\: \hline or \cline{col1-col2} \cline{col3-col4} ...
Sequence name & Y & U & V               
  \\
  \hline
  Train\_le                & --0.1\%  &  --0.1\%  & --0.1\% \\
  SkateBoardingTrick\_le   & --0.4\%  &  --1.1\%  & --0.8\% \\
  SkateboardInLot          & --1.6\%  &  --1.7\%  & --1.9\% \\
  ChairLift                & --3.0\%  &  --4.0\%  & --3.6\% \\
  KiteFlite                & --0.1\%  &  --0.2\%  & --0.1\% \\
  Harbor                   & 0.0\%    &  0.0\%    & 0.1\%   \\
  PoleVault\_le            & --0.2\%  &  --0.1\%  & --0.2\% \\
  AerialCity               & --2.6\%  &  --2.5\%  & --2.7\% \\
  DrivingInCity            & --0.9\%  &  --1.6\%  & --0.8\% \\
  DrivingInCountry         & --3.4\%  &  --3.4\%  & --4.5\% \\
  average                  & --1.2\%  &  --1.5\%  & --1.5\% \\
  \hline
\end{tabular}
\end{center}
\end{table}

\begin{table}[t]
\begin{center}
\caption{The performance in LDP case} \label{tab:LDP}
\begin{tabular}{|c|c|c|c|}
  \hline
  % after \\: \hline or \cline{col1-col2} \cline{col3-col4} ...
Sequence name & Y & U & V               
  \\
  \hline
  Train\_le                & --0.1\%  &  --0.1\%  & --0.1\% \\
  SkateBoardingTrick\_le   & --0.3\%  &  --0.7\%  & --0.6\% \\
  SkateboardInLot          & --1.8\%  &  --1.9\%  & --0.7\% \\
  ChairLift                & --2.9\%  &  --3.8\%  & --3.0\% \\
  KiteFlite                & --0.1\%  &  --0.2\%  & --0.3\% \\
  Harbor                   & 0.0\%    &  0.0\%    & 0.2\%   \\
  PoleVault\_le            & --0.1\%  &  0.2\%    & --0.2\% \\
  AerialCity               & --2.5\%  &  --2.5\%  & --2.2\% \\
  DrivingInCity            & --0.7\%  &  --1.1\%  & --1.1\% \\
  DrivingInCountry         & --3.3\%  &  --3.2\%  & --3.8\% \\
  average                  & --1.2\%  &  --1.3\%  & --1.2\% \\
  \hline
\end{tabular}
\end{center}
\end{table}

Except for the average and maximum bitrate reduction, we can also see that the proposed algorithm can lead to consistently better R-D performance for all the test sequences even if the RDO based selection between the proposed reference frame and the original reference frame is not used in the proposed framework.
This can obviously demonstrate that the reference frame in the proposed framework can always lead to better or equivalent compression results compared with that in the original framework.
However, we can also see that the performance improvement may vary due to the differences of the characteristics of various sequences.
For the sequences with large motion in the face boundary such as the sequence DrivingInCountry, the situation where the MC cross the face boundary will be quite a lot, thus the proposed algorithm can lead to significant bitrate reduction.
On the contrary, for the sequences with almost zero motion in the face boundary such as the sequence Harbor, the situation where the MC cross the face boundary will be very rare, thus the proposed algorithm cannot provide an obvious performance improvement.

\begin{figure}[t]
\begin{minipage}[b]{0.98\linewidth}
  \centering
  \centerline{\includegraphics[width=7.0cm]{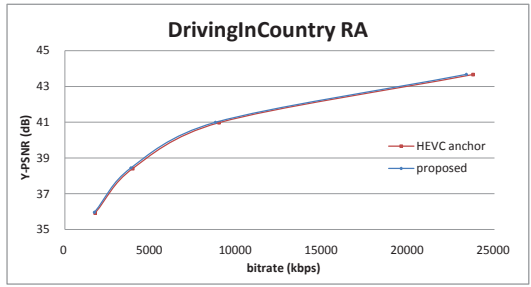}}
  \centerline{(a) RA}\medskip
\end{minipage}
\begin{minipage}[b]{0.98\linewidth}
  \centering
  \centerline{\includegraphics[width=7.0cm]{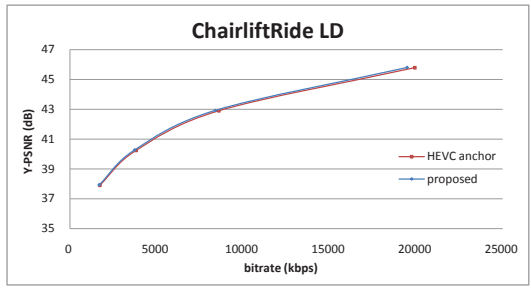}}
  \centerline{(b) LD}\medskip
\end{minipage}
\begin{minipage}[b]{0.98\linewidth}
  \centering
  \centerline{\includegraphics[width=7.0cm]{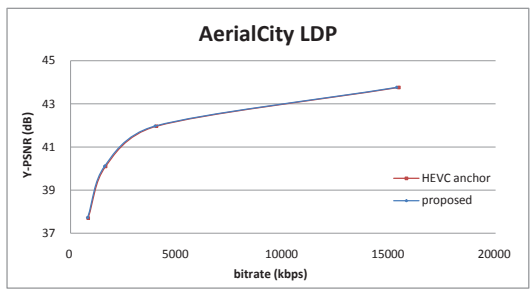}}
  \centerline{(c) LDP}\medskip
\end{minipage}
\caption{Typical R-D curves}
\label{fig:curve}
\end{figure}

Some typical R-D curves in various test conditions with different test sequences are shown in Fig. \ref{fig:curve}.
The R-D curves also demonstrate that the proposed algorithm can lead to some performance improvement compared with HEVC anchor.
Besides, from these typical R-D curves, we can also see that the proposed algorithm can lead to similar performance improvement for both high bitrate and low bitrate.

\section{Conclusion}
\label{sec:conclusion}
In this paper, we first point out the existence and influences of the very serious texture discontinuities in the face boundary in the polyhedron projection.
Then we propose to fill the corresponding neighboring faces in the suitable positions as the extension of the current face to keep approximated texture continuity.
After that, a co-projection-plane based 3-D padding method is proposed to project the reference pixels in the neighboring face to the current face to guarantee exact texture continuity.
The proposed scheme is implemented in the reference software of High Efficiency Video Coding. 
Compared with the existing method in the High Efficiency Video Coding reference software, the proposed algorithm can bring averagely $1.1\%$ and maximum $3.4\%$ bitrate savings in different test conditions.
The experimental results obviously demonstrate that the texture discontinuity in the face boundary can be well handled by the proposed algorithm.

% -------------------------------------------------------------------------
\bibliographystyle{IEEEbib}
\bibliography{icme2017template}

\end{document}